\documentclass[utf8]{frontiersSCNS} 

\usepackage{url,hyperref,microtype,subcaption}
\usepackage[onehalfspacing]{setspace}

\usepackage{ulem}

\def\keyFont{\fontsize{8}{11}\helveticabold }
\def\firstAuthorLast{V. Popov {et~al.}} 
\def\Authors{Vladislav Popov$^{1}$, Mikhail Odit$^{1}$, Jean-Baptiste Gros$^{1}$, Vladimir Lenets$^{1}$, Akira Kumagai$^{3}$, Mathias Fink$^{2}$, Kotaro Enomoto  $^{3}$, and  Geoffroy Lerosey$^{1*}$} 
\def\Address{$^{1}$Greenerwave, 6 rue Jean Calvin, 75005 Paris, France \\
$^{2}$Institut Langevin, 1 rue Jussieu, 75005 Paris, France \\
$^{3}$AGC Inc.,1-1, Suehiro-cho, Tsurumi-Ku, Yokohama-shi, Kanagawa 230-0045, Japan 
}
\def\corrAuthor{Geoffroy Lerosey}

\def\corrEmail{geoffroy.lerosey@greenerwave.com}

\begin{document}
\onecolumn
\firstpage{1}

\title[]{
Experimental demonstration of a mmWave passive access point extender based on a binary reconfigurable intelligent surface
} 

\author[\firstAuthorLast ]{\Authors} 
\address{} 
\correspondance{} 

\extraAuth{}

\maketitle

\begin{abstract}

\section{}
As data rates demands are exploding, 5G will soon rely on mmWaves that offer much higher bandwidths. Yet at these frequencies, attenuation and diffraction of waves require point to point communications with beamforming base stations that are complex and power greedy. Furthermore, since any obstacle at these frequencies completely blocks the waves, the networks must be extremely dense, resulting in dramatic increase of its cost. One way to avoid this problem is to redirect beams coming from base stations at many locations with Reconfigurable Intelligent Surfaces, in order to increase their coverage even in cluttered environments. Here we describe and experimentally demonstrate a binary tunable metasurface operating at 28~GHz, based on standard PCB and off the shelves PIN diodes. We show that it can be used as a Reconfigurable Intelligent Surface that beamforms an incoming plane wave at a given angle to one or several outgoing plane waves at angles reconfigurable in real time. Most importantly we use this 20~cm$\times$20~cm reconfigurable Intelligent Surface alongside software defined radio and up/down converters at 28~GHz, and demonstrate a wireless link between an emitter and a receiver 10~meters away, in a non line of sight configuration, hence proving the validity of the approach.

\tiny
 \keyFont{ \section{Keywords:} Reconfigurable intelligent surface, smart walls, metasurface, RIS, MIMO, mmWave, 5G, digital beamforming } 
\end{abstract}

\section{Introduction}

From its inception with Marconi and Hertz one century ago \citep{hertz1893,Marconi1909}, the field of wireless communications has seen tremendous technological revolutions, which have resulted in impressive new possibilities, and ultimately to their global usage of nowadays and to the growing bandwidth needs we see. For instance, if the concept of beamforming has been known almost from the start \citep{Brau1909}, it has only become a reality in wireless systems recently, owing to the development of cost effective and compact radiofrequency integrated circuits. Similarly, the impressive usage allowed by 4G and now 5G, with real time video meetings for instance, would not have been possible without processors incredible breakthroughs of the past decades.

In physics, the past 20 years have been marked by the era of new man-made materials with exotic and non-naturally existing properties: photonic crystals \citep{Joannopoulos2011}, metamaterials \citep{Capolino2017} and metasurfaces \citep{Holloway2012}. Meanwhile, impressive approaches were proposed aiming at controlling waves in very complex and scattering media, from acoustics to optics \citep{Fink1997,Lerosey2007,Mosk2012}. Working at the interface between these two fields of research persuaded some of us, a little less than 10 years ago, that the future of wireless communications was in electronically reconfigurable materials and smart electromagnetic environments for superior wave control and optimal data rates.

In a seminal set of papers \citep{Kaina:14,Kaina2014,Dupre2015}, we proposed to cover part of the walls of an office with real-time electronically tunable metasurfaces that control in a passive way the reflections of the electromagnetic waves emitted by a remote isotropic source. We showed that such approach could improve 10 fold the signal received by a receiver, or on the contrary completely null it, with close to zero electric consumption and very inexpensive hardware. This was the first proposal of the now called Reconfigurable Intelligent Surfaces concept (RIS).

The concept went quite unnoticed for some years, while the wireless community’s attention was more on 5G related technologies such as Massive-MIMO \citep{Marzetta2013,Lu2014,Bjornson2015}. Then the topic started to gain momentum two years ago, as it was naturally taken over by the wireless communications community, making now RIS one of the hottest topics in the field. Meanwhile it gave us some time to develop the technology at Greenerwave \citep{Greenerwave}, and to foresee where it could first have practical applications. For standardization and practical reasons, we believe the first real product based on RIS could be a passive access point extender for mmWave, and this is what we propose in this paper.

We first build on a recent paper, where we proposed to control radiofrequency waves at 28~GHz using a binary reconfigurable metasurface as a RIS \citep{RIS_Greenerwave}, and describe a 20~cm$\times$20~cm FPGA controlled RIS that can act as an access point extender for 5G mmWave. In a third part we show how such RIS can be used to beamform on a given user, and even on several users, simply relying on analytical formulas. 
In a final part, we demonstrate how our RIS can perform real data link optimization, using software defined radio for the baseband and up/down converters at mmWave in the 5G band (28~GHz). Most importantly, we show that our tunable metasurface can be used to guarantee a data link in a non-line-of-sight situation where mmWaves are conveyed between a corridor and a room, proving the viability of RIS as passive access point extenders for 5G networks.

\section{RIS Design}
\begin{figure}[h!]
\begin{center}
\includegraphics[width=0.65\columnwidth]{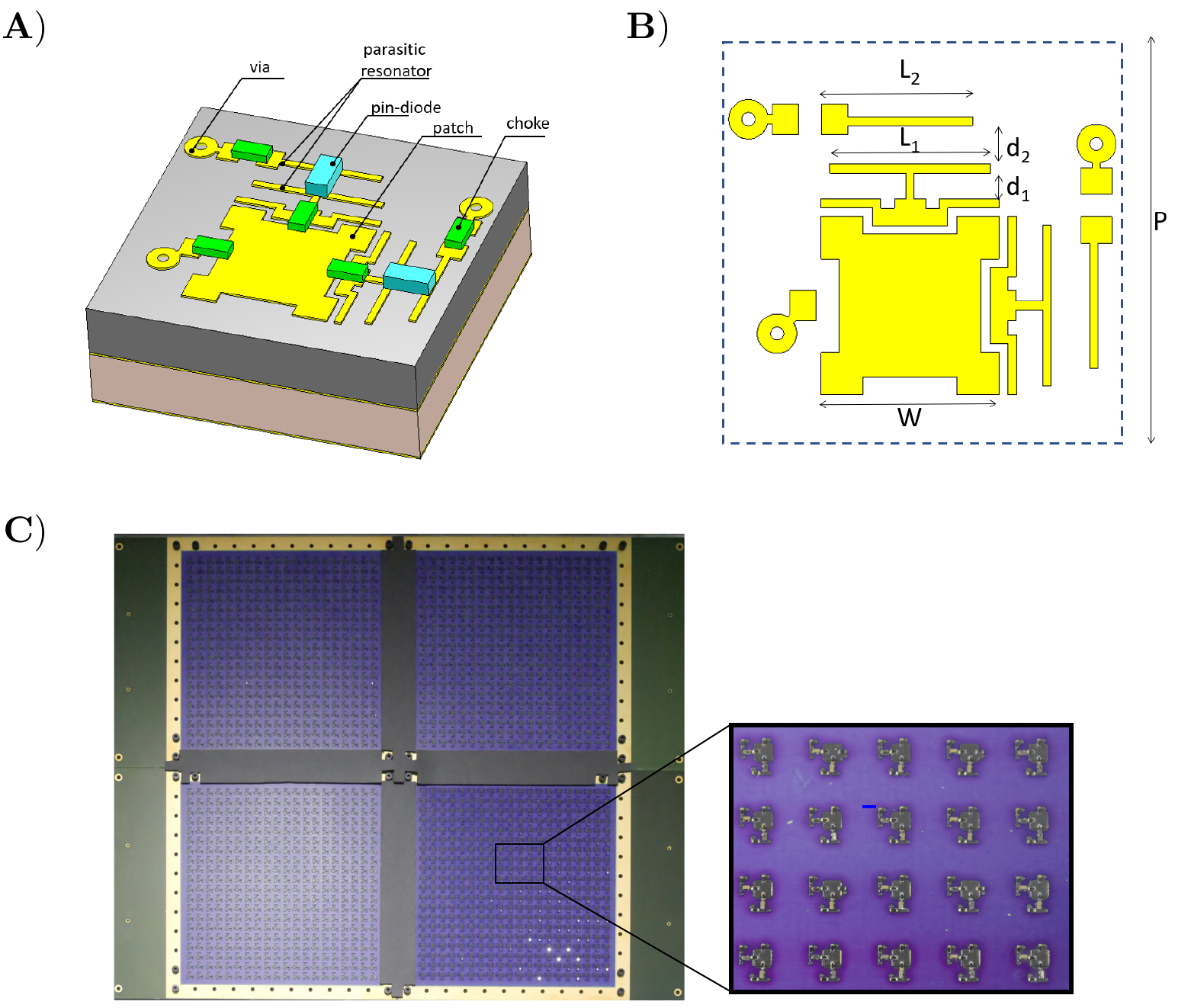}
\end{center}
\caption{The perspective view \textbf{(A)} and layout \textbf{(B)} of the RIS unit cell. \textbf{(C)} Photo of the fabricated $20\mathrm{cm} \times 20\mathrm{cm}$ RIS composed of 4 $10\mathrm{cm} \times 10\mathrm{cm}$ reflect arrays, controlled by one FPGA board. The total number of unit cells is 1600, each with independently control of 2 polarizations.}\label{fig:1}. 
\end{figure}

In this work we introduce a reconfigurable intelligent surface (RIS) based on the original design of a reflecting metasurface first described in \citep{RIS_Greenerwave}. The metasurface is comprised of a set of periodically arranged unit cells with a $\lambda/2$ spacing. 
The design of a single unit cell is shown on figure~\ref{fig:1}A. 
The detailed unit cell description is provided in the above-cited publication by the authors. 
The unit is designed to control the phase of the reflected field in order to provide a close to the $\pi$ phase difference  between 2 different states of the electronically controlled element. 
The control of the reflection field for each polarization is provided by the mutual coupling of the resonances of the main patch resonator and parasitic resonator placed nearby \citep{Kaina:14}.
The switching between resonance-states is performed by the control of the impedance of the PIN diode (MACOM MADP-000907-14020) which is RF-isolated from the biasing voltage through AVX L0201 SMD Inductors (RF chokes). 
The PIN diodes and chokes are shown correspondingly with blue and green colors in figure~\ref{fig:1}A.
The biasing voltage is applied to the PIN diodes through the vias.
The following dimensions of the elements of the pixels, see figure~\ref{fig:1}B, are used in the design:
$L_1=1.8$~mm$,L_2=1.7$~mm$,d_1=0.3$~mm$,d_2=0.44$~mm$,W=2$~mm$, P=5$~mm. 
The PCB is built of two substrates: Meteorwave 8000~\citep{meteorwave} and FR-4.
The Meteorwave 8000 substrate  ($\varepsilon=$ 3.3 and $tan(\delta)=$ 0.0016 at 10~GHz) is used as a top dielectric layer to hold resonating elements.
The FR-4 substrate  ($\varepsilon=$ 4.3 and $tan(\delta)=$ 0.025 at 10~GHz) is used as a bottom layer which is intended to host electronic circuits for the biasing voltage of the PIN diodes.

The $20\times20$ dual polarization unit cells are placed each $\lambda/2$ (at 30~GHz) on a 10$\times$10~cm PCB board. Several similar boards can be used in order to increase the total RIS aperture size and consequently the directivity. Here we assemble RIS from four joined together PCBs in order to have a $20$~cm~$\times$~$20$~cm aperture (see figure~\ref{fig:1}C). 
Characteristics of the built RIS are summarized in the table \ref{tab1}.

The RIS  is controlled by means of an FPGA control board, which is designed to independently switch each of the 3200 PIN diodes of the RIS. 
The power consumption of the control board having 4 connected metasurfaces starts from a few watts when the PIN diodes are in off state and reaches 16 W when all the PIN diodes are switched on. 
The average power consumption (implying mixed states of the diodes) is typically below 6 W (see table \ref{tab1}). 
The switching rate of the RIS state reaches 100 kHz if the pattern of the PIN diodes states is stored in local memory of the board.

\section{Beamforming with RIS}

In this section we consider algorithms for controlling a RIS.
Depending on the relative positions of a transmitting (Tx) antenna, receiving (Rx) antenna and a RIS, one can identify two different modes of RIS operation: near-to-far field and far-to-far field. 
In the first scenario, one of the antennas (Tx for instance) is placed in the near-field of a RIS, i.e. at a distance when the incident wave cannot be considered as a plane wave. 
In this mode a RIS and the Tx antenna resemble a common reflect array antenna. 
The signal emitted by the Tx antenna is reflected by the RIS and focused at an Rx antenna, which is placed in the far-field region. 
In the second case a RIS operates in the far-to-far field configuration, implying that both Rx and Tx antennas are placed in the far-field of a RIS.
This scenario is supposed to be more practical and more common in real-life 5G communication systems, where the RIS can be used as a passive or active access point extender that redirects a signal from a base station towards a user device.
In what follows we focus on the developed $20$~cm~$\times$~$20$~cm RIS operating in the far-to-far field mode.

\subsection{Physical model of RIS}
When an electromagnetic wave impinges the RIS, the reflected wave will depend on its configuration.
In order to increase the signal-to-noise ratio at a receiver, the RIS must form a beam in its direction.
A simple analytical model considering the RIS as an array of  elements with $\cos(\theta)^q$ radiation pattern~\citep{yang2016programmable,RIS_Greenerwave} can be used to find its configuration for given positions of Tx and Rx antennas.
According to this model the electric field of the reflected wave in the direction $(\theta_{Rx}, \varphi_{Rx})$ of the Rx antenna is calculated as follows
\begin{equation}
    \label{eq:reflected_wave}
    E_r(\theta_{Rx},\varphi_{Rx}) =
    \cos(\theta_{Rx})^q \cos(\theta_{Tx})^q 
    \sum_{n,m = 1}^N  \Gamma_{nm} E_i(x_n,y_m)e^{-jk\sin(\theta_{Rx})[x_n\cos(\varphi_{Rx})+y_m\sin(\varphi_{Rx})]},
\end{equation} 
where $\theta$ and $\varphi$ are respectively elevation and azimuth angles of the spherical coordinate system having its origin in the center of the RIS, $\theta_{Tx}$ and $\theta_{Rx}$ are the elevation angles towards Tx and Rx antennas, respectively.
The reflection coefficient from the $nm^\textup{th}$ pixel is denoted by $\Gamma_{nm}$ and  
the electric field of the impinging wave at the position of the $nm^\textup{th}$ pixel is $E_i(x_n,y_m)$. 
The reflection coefficient can be expressed in terms of the configuration of the RIS as $\Gamma_{nm} = \Gamma_{\textup{ON}}C_{nm} + \Gamma_{\textup{OFF}}(1-C_{nm})$, where $C_{nm}$ equals one in ON-state and zero in OFF-state of the $nm^\textup{th}$ pixel. 

\subsection{Redirecting signal towards user device}

In order to maximize the received signal, the absolute value of the right-hand side of Eq.~\eqref{eq:reflected_wave} is maximized with respect to the configuration of the RIS.
To that end, a simple iterative optimization algorithm can be applied following~\citep{mosk2008}.
In the first step all pixels are put in OFF-state.
Then, the first pixel is turned ON ($C_{11}=1$) and is kept in ON-state, if the goal function increases. 
Otherwise, the pixel is turned back to OFF-state.
The same algorithm is applied to all pixels one-by-one.
The number of iterations is limited to the total number of pixels constituting the RIS and equals $N^2$.
In comparison to a purely analytical approach of phase gradient typically used for reflectarrays~\citep{pozar1997design,RIS_Greenerwave} and transmitarrays~\citep{clemente2013wideband}, the optimization algorithm allows one to achieve sightly better directivity and level of side lobes.

Two examples of applying the presented algorithm are shown in Figure~\ref{fig:2}, when the incident wave is a plane wave impinging the RIS from ($-45^\circ$ azimuth, $0^\circ$ elevation). 
Panels (A) and (B) demonstrate color-maps of the directivity calculated by means of the model, when the RIS is configured to redirect the incident wave towards a receiver at ($0^\circ$ azimuth, $-30^\circ$ elevation) and ($20^\circ$ azimuth, $20^\circ$ elevation).
The RIS configurations corresponding to the far-field radiation patterns appear right below, in panels (D) and (E). 
The white cross in the middle represents the separation of 2 cm between the four metasurfaces constituting the RIS.
In the panels (A) and (B) one can identify two beams: the desired beam (highlighted by a white circle) and the spurious beam (highlighted by a red circle). 

The appearance of the spurious beam is a consequence of the binary phase control provided by the RIS in the far-to-far field operation mode (both incident and reflected waves have planar wavefront)~\citep{cui2014coding,RIS_Greenerwave}.
The spurious beam disappears when one of antennas is in the near-field of the RIS (the incident wave is rather a spherical than a plane wave).
Curiously, the spurious beam does not significantly reduce the directivity in comparison to the equivalent configuration of the RIS operating in the near-to-far field mode as it was demonstrated in our earlier work~\citep{RIS_Greenerwave}.
Furthermore, the two modes result in similar directivities for the same directions of the desired beam.
This fact allows one to approximately estimate the directivity of the RIS in the far-to-far field mode (that can be challenging to measure directly in the experiment) by measuring the radiation pattern of the RIS created in the near-to-far field mode.
This is procedure was used in order to estimate the directivity parameter in Table~\ref{tab1} for our RIS operating in the far-to-far field mode.

\subsection{Serving multiple users at the same time}

The same RIS can be used serve multiple users at the same time.
First of all, the RIS developed in this work allows one to independently control both orthogonal polarizations of an impinging wave.
It means that two independent bit-streams can be transferred at the same time when attributed to different polarizations.  

On the other hand, a RIS can create multiple beams in independent directions.
In this case frequency-division multiplexing can be used for transmitting two independent data streams simultaneously and without the need to switch a RIS between two users.
For instance, an optimization procedure based on the one described above can be used to put the RIS in a double-beam configuration, i.e. to redirect a signal from a base-station towards two receivers at different locations.
In the first step of this procedure a RIS is put in a single-beam configuration towards the first receiver.
The second step is to select a random pixel and put it in the state opposite to the current one. 
The state of the selected pixel is kept, if the amplitude of the signal at the second receiver increases. 
Otherwise, the pixel is flipped back.
The second step is repeated until the amplitudes of the signal at the first and second receivers are equal.
After that, the optimization procedure is completed and final configuration set-ups a RIS to create two beams.

\begin{figure}[h!]
\begin{center}
\includegraphics[width=0.9\linewidth]{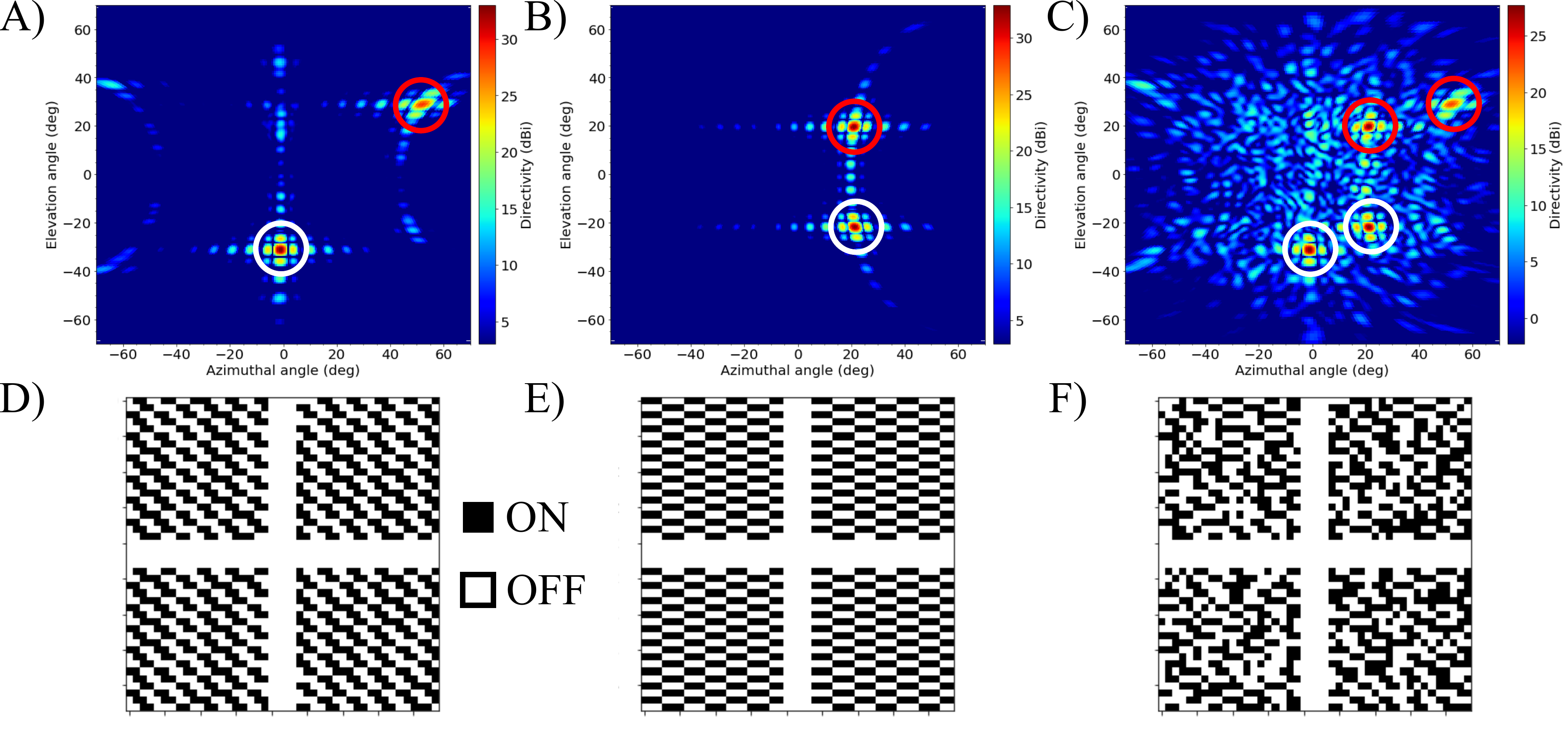}
\end{center}
\caption{(A) -- (C) Numerically calculated far-field radiation patterns created by the RIS illuminated by a plane wave at $-45^\circ$ azimuth incidence and set in a single-beam (A,B) and a double-beam (C) configurations. 
(D)--(F) Corresponding binary configurations. The white cross represents the separation of 2 cm between the four metasurfaces constituting the RIS.}\label{fig:2}. 
\end{figure}

Figure~\ref{fig:2}(C) demonstrates a radiation pattern from a RIS in a double-beam configuration, when the impinging wave arrives from the direction ($-45^\circ$ azimuth, $0^\circ$ elevation) and is redirected towards two independent receivers at ($0^\circ$ azimuth, $-30^\circ$ elevation) and ($20^\circ$ azimuth, $20^\circ$ elevation). 
The corresponding RIS configuration is shown in Figures~\ref{fig:2} (F).
Similarly to the single-beam scenarios, there are two desired beams and two spurious beams highlighted by  white and red circles, respectively.

Two beams is not a limit for a RIS and more beams can be controlled simultaneously thus allowing a single base station to serve multiple users at the same time in a NLOS situation.
However, multiple-beam control comes at the price of a reduced gain in a given direction, what should be taken into account when calculating the link budget.

\section{Indoor wireless communication with RIS}

At mmWaves a signal from a base station experiences significant fading due to propagation loss and shadowing by buildings, landscape or indoor walls.
This problem can be solved by installing additional base stations or access point extenders (passive or active) between existing base stations. 
In this section we demonstrate experimentally the integration of our RIS in an indoor wireless communication system.
This experiment shows a great potential of RIS to provide a robust broadband connectivity even in NLOS situations.

\begin{figure}[h!]
\begin{center}
\includegraphics[width=0.9\columnwidth]{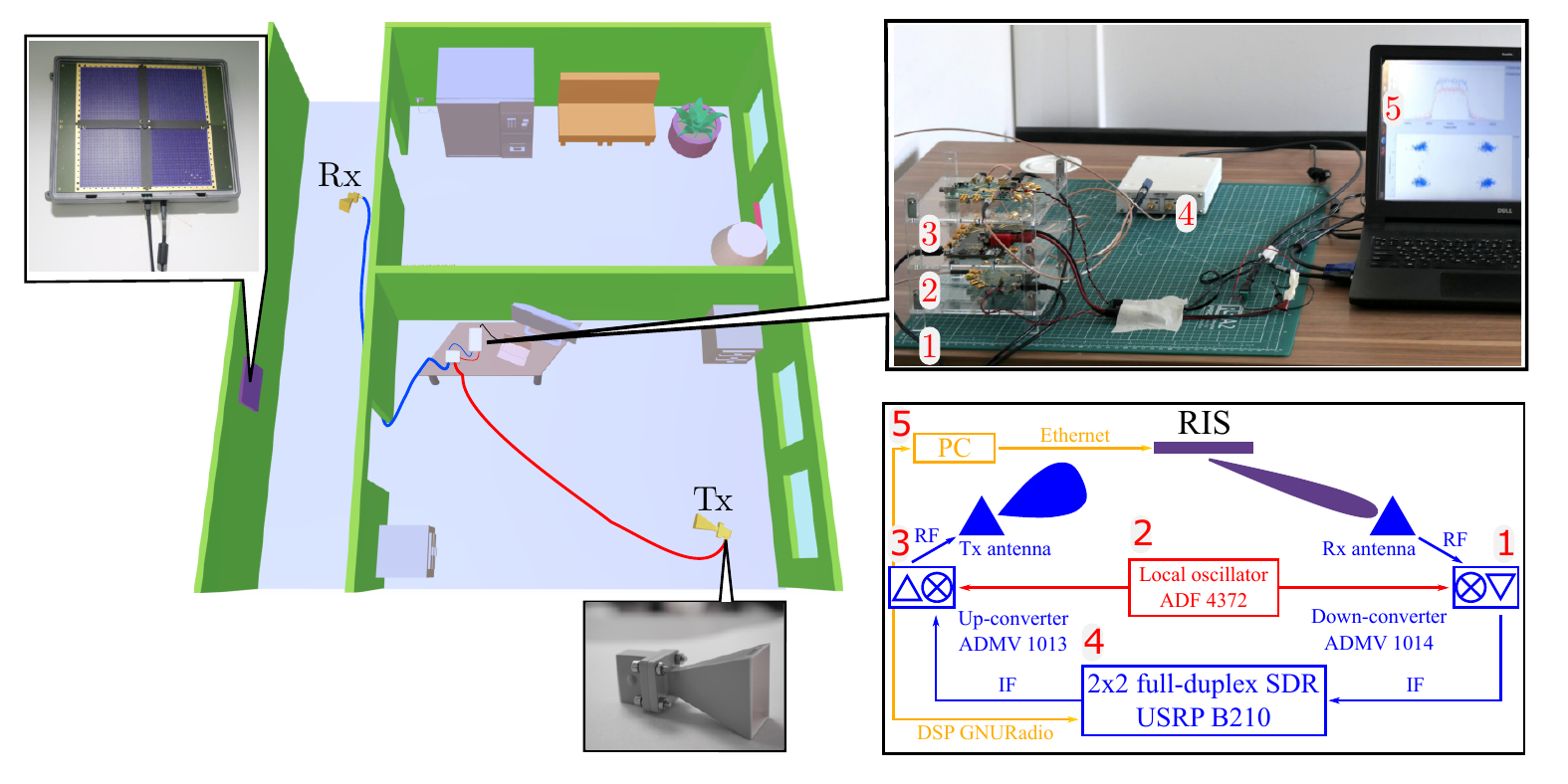}
\end{center}
\caption{
Left: A 3D illustration of the experimental setup used to demonstrate RIS-aided indoor wireless communication in a NLOS situation. 
Right: A photography (top panel) and a schematic (bottom panel) of the communication module represented by a down-converter \textbf{1}, a local oscillator \textbf{2}, an up-converter \textbf{3}, a SDR \textbf{4}, a PC \textbf{5}.
}
\label{fig:3}
\end{figure}

To demonstrate operation of the RIS as a passive access point extender we set up an indoor wireless communication system as illustrated by Figure~\ref{fig:3}. 
Here the RIS is placed on a wall of a corridor connecting offices. The Rx and Tx are WR-34 Pyramidal Horn Antennas (Rx) emulating correspondingly an end user terminal (i.e. mobile phone or laptop) and a 5G base station antenna. 
The distance from the RIS to the Rx antenna is $2$m, the azimuth angle is $52.5^\circ$ and the elevation angle is $-20^\circ$.
A Tx horn antenna is placed inside an office room providing non-line-of-sight (NLOS) configuration with the Rx antenna.
The distance from the RIS to the Rx (Tx) antenna is $2$m ($5.5$m), with respect to the normal to the RIS the azimuth angle is $52.5^\circ$ ($-35^\circ$) and the elevation angle is $-20^\circ$ ($0^\circ$).

We start by using a Rohde\&Schwartz ZVA20 vector network analyzer (VNA) to 
characterize the channel when the RIS is put in a beamforming configuration or is switched off. 
In the beamforming configuration the RIS creates a LOS between the Tx and Rx antennas.
However, in order to efficiently redirect the wave radiated by the Tx antenna towards the Rx antenna, it is necessary to know their precise positions with respect to the RIS.
In the given experimental setup we first estimate these locations approximately and then use a feedback from the VNA ($S_{21}$ parameter) to scan the space in the vicinity of the estimated location antenna positions.
In the scanning procedure the RIS is configured according to estimated positions of the Rx and Tx antennas at given iteration. 
These positions are used as input parameters for the beamforming algorithm described in the previous section. 
When the $S_{21}$ parameter is maximized, precise positions of the Tx and Rx antennas are found and the RIS redirects the wave radiated by the Tx antenna towards the Rx antenna establishing the LOS.

\begin{figure}[h!]
\begin{center}
\includegraphics[width=0.9\linewidth]{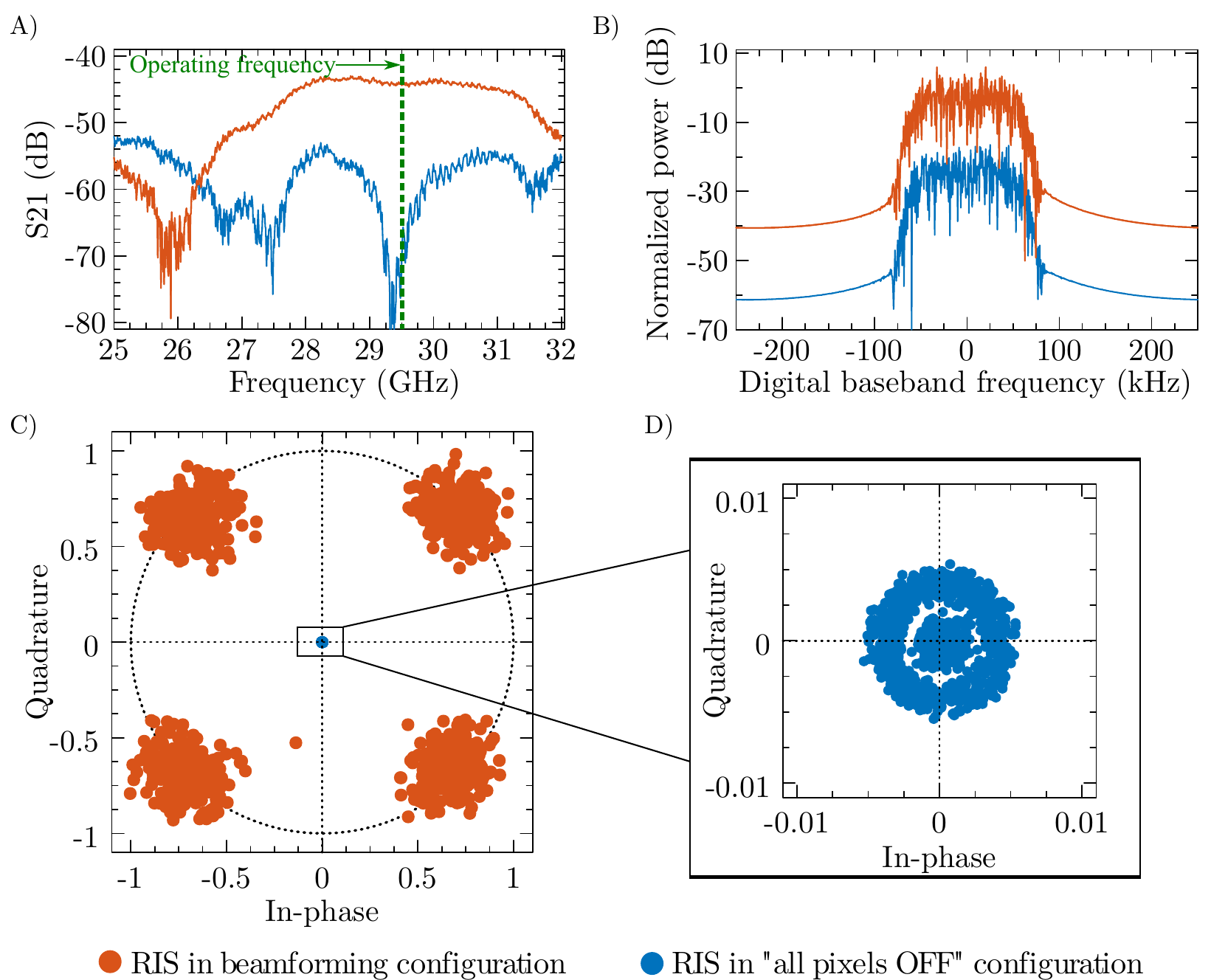}
\end{center}
\caption{
(A) S21 parameter measured with a VNA for channel characterization in two configurations of the RIS: all pixels off and beamforming configuration.
(B) The signal received by the SDR and put through an RRC filter, when the RIS is off and on.
(C) Constellation diagram after timing recovery, equalization and phase and frequency offset compensation, when the RIS is off and on.
(D) Close-up of the constellation diagram when the RIS is off.
}
\label{fig:4}
\end{figure}

Figure~\ref{fig:4}(A) compares the path-loss (measured with the VNA) between the Tx and Rx antennas, when the RIS is switched off (the blue curve) and when the RIS is on creating the LOS (the orange curve).
It is seen that the RIS allows one to increase the signal power by 30~dB and maintain it at almost constant level in a wide frequency range.
The instantaneous bandwidth constitutes more than $3$ GHz with respect to 29.5 GHz central frequency.
On the other hand, when the RIS is switched off the signal level drops below the noise level at $29.5$~GHz and significantly varies with respect to the frequency. 



After the channel is characterized, the VNA is substituted by a wireless communication system shown in Figure~\ref{fig:3} on the right panels.
Unfortunately, we did not have an access to a real base station and had to built our own setup.
In this setup a full duplex software-defined-radio USRP B210 by Ettus Research (SDR) is used in the communication system as a transmitter and a receiver of a modulated signal at an intermediate frequency (IF).
The SDR generates a waveform at IF in the Tx channel and provides in-phase and quadrature-phase (IQ) received signal in the digital baseband for further signal processing. 
The RF front-end of the communication system is represented by an up-converter, a down-converter and a local oscillator (LO) by Analog Devices (ADMV 1013, ADMV 1014 and ADF 4372, respectively).
Up- and down-converters multiply the LO frequency by four.
A PC controls the RIS via an Ethernet interface and performs digital baseband signal processing of the signal received by the SDR.
The signal processing is done with the help of an open software platform GNURadio.
A photography of the setup and its schematic are shown on the top right of the Figure~\ref{fig:3}.

To transmit at the RF frequency of 29.5 GHz, the LO is set at 7 GHz and IF generated by the SDR is set to 1.5 GHz.
As a data stream was chosen a large array of random numbers, QPSK modulation followed by root-raised-cosine (RRC) filter creates the transmitted waveform.
On the receiver chain, the signal received by the SDR is first put through the same RRC filter, after that timing recovery, equalization and compensation for phase and frequency offset is performed. 
Finally, the received signal is demodulated.
It is important to note that even though the Tx and Rx chain of the SDR uses the same clock, one should compensate for the possible frequency offset after the up- and down-conversion processes.

Figure~\ref{fig:4}(B) shows normalized power of the received signal at the digital baseband, when the RIS is switched off (the blue curve) and when the RIS is put in the beamforming configuration (the orange curve) to redirect the impinging wave from the Tx antenna towards the Rx antenna.
The beamforming by the RIS allows one to enhance the level of the received signal by more than 25 dB and the four symbols of the QPSK modulation are well-distinguished. 
When the RIS is off, the receiver cannot even lock the constellation as demonstrated by Figures~\ref{fig:4}(C) and (D).

The results of this section clearly demonstrate how our RIS allows one to restore a data transmission between two antennas in the NLOS situation and establish a solid mmWave communication channel in a complex indoor environment.
Furthermore, additional functionality can be added, when a single base station can serve multiple users simultaneously with the help of a RIS.
In this case frequency division, code division or time division multiple access techniques can be used~\cite{miao2016fundamentals}.


\section{Conclusion}

To conclude, in this paper we have presented a design of a 20~cm$\times$20~cm  binary reconfigurable intelligent surface (RIS) operating at 28~GHz frequency range. 
The RIS consists of 1600 independently controlled unit cells. Each of them interact with both horizontal and vertical polarisations of the incident field providing reflection of the wave with variable phase and dissipation less than -3dB. 
We show that it is possible to perform analytical beamforming with the RIS imposing proper phase distributions to the RIS elements and recovering the communication channel between two antennas positioned in the non-line-of-sight configuration. 
When the RIS is switched off the signal doesn't propagate between antennas. When the RIS is on the signal level is increased 30 dB. Using software defined radio and up/down converters, we finally prove that such RIS can restore a data link in a non line of sight configuration, in a typical office setup.  

The experiments presented in this paper prove the viability of the concept of passive access point extenders at mmWave, based on tunable metaurfaces used as Reconfigurable Intelligent Surfaces.

\begin{table}[tb]
    \centering
        \caption{RIS specifications table.}
    \label{tab1}
    \begin{tabular}{|c|c|}
    \hline
    Dimensions & $20$ cm $\times$ $20$ cm\\ \hline
    Number of unit cells  & 1600 \\ \hline
    Polarization & dual linear \\ \hline    
    Element's spacing & $\lambda/2$ \\  \hline
    Operating frequency range  & $[26 - 31]$ GHz\\  \hline
    Total bandwidth  & 5~GHz\\  \hline
    Reflection loss within bandwidth  & $<3$~dB\\  \hline
    Instantaneous bandwidth & 3~GHz\\ \hline
    Directivity  & $30$ dBi\\  \hline
    Scan range (elevation, azimuth) & $\pm 60^\circ$\\ \hline
    Beam switch rate & 100 kHz\\   \hline
    Power consumption (min / average / max) & 3/6/16 W\\   \hline
    \end{tabular}

\end{table}


\section*{Conflict of Interest Statement}
 The authors declare that the research was conducted in the absence of any commercial or financial relationships that could be construed as a potential conflict of interest.

\section*{Author Contributions}
All the authors contributed equally to the paper.


\section*{Acknowledgments}
Greenerwave acknowledges the support from the European Commission through the H2020 Project through the RISE-6G, HEXA-X under Grant 101015956.



\bibliographystyle{frontiersinSCNS_ENG_HUMS} 
\bibliography{main}

\end{document}